\documentclass[aps,pre,twocolumn,groupedaddress,showpacs]{revtex4}
\usepackage{graphicx}
\usepackage{amssymb}

\begin{document}
%\preprint{}
\title{Topological persistence and dynamical heterogeneities near jamming}

\author{A.R. Abate and D.J. Durian}
%\thanks{}
\affiliation{Department of Physics \& Astronomy, University of
Pennsylvania, Philadelphia, PA. 19104-6396}


\date{\today}

\begin{abstract}
We introduce topological methods for quantifying spatially heterogeneous dynamics, and use these tools to analyze particle-tracking data for a quasi-two-dimensional granular system of air-fluidized beads on approach to jamming.  In particular we define two overlap order parameters, which quantify the correlation between particle configurations at different times, based on a Voronoi construction and the persistence in the resulting cells and nearest neighbors. Temporal fluctuations in the decay of the persistent area and bond order parameters define two alternative dynamic four-point susceptibilities, $\chi_A(\tau)$ and $\chi_B(\tau)$, well-suited for characterizing spatially-heterogeneous dynamics.  These are analogous to the standard four-point dynamic susceptibility $\chi_4(l,\tau)$, but where the space-dependence is fixed uniquely by topology rather than by discretionary choice of cutoff function.  While these three susceptibilities yield characteristic time scales that are somewhat different, they give domain sizes for the dynamical heterogeneities that are in good agreement and that diverge on approach to jamming.
\end{abstract}

\pacs{45.70.–n, 47.55.Lm, 64.70.Pf, 02.40.Pc}

% 47.55.Lm Fluidized beds
% 45.70.-n Granular systems
% 02.40.Pc topology
% 64.70.Pf Glass transitions
%
%
%
%

%\keywords{}

\maketitle

%\section{}
%\subsection{}
%\subsubsection{}

% If in two-column mode, this environment will change to single-column
% format so that long equations can be displayed. Use
%\begin{widetext}
%\end{widetext}

%=========================================================================================

\section{Introduction}

The dynamics in both thermal and driven systems become increasingly heterogeneous \cite{EdigerSHDinSCL, GlotzerJNCS00, CipellettiRamos05} on approach to jamming, where a control parameter such as temperature, packing fraction, or driving rate is varied to the point that all rearrangements cease~\cite{CatesPRL98, LiuNagelBOOK}.  Such spatially-heterogeneous dynamics (SHD) may take the form of intermittent strings or swirls of neighboring particles that follow one another in correlated fashion from one packing configuration to another.  Observations of SHD have been reported for coarsening \cite{DJDsci91,TRConFoam} and sheared \cite{AnthonyPRL95} foams, for dense colloidal suspensions \cite{MarcusPRE1999ColloidalStrings, KegelScience2000ColloidalStrings, WeeksScienceColloidalStrings}, and for driven granular systems \cite{PouliquenPRL03,DauchotPRLSubDiff2005}.  Quantification of SHD may be accomplished by a dynamic four-point susceptibility, $\chi_4(l,\tau)$, defined in terms of density correlations across intervals of both space and time \cite{LacevicJCP2003X4}.  Bounds on $\chi_4(l,\tau)$ have been inferred from standard measurements of dielectric susceptibility for glass-forming liquids and of dynamic light scattering for dense colloidal suspensions \cite{BerthierScienceX4DynLength}.  Direct measurements of $\chi_4(l,\tau)$ have been reported for quasi-two-dimensional granular systems, where particle motion was excited by shear \cite{DauchotSHDPRL2005} and by a fluidizing upflow of air \cite{KeysAbateNP07}.  In the former, $\chi_4(l,\tau)$ was measured vs space and time at a fixed packing fraction; in the latter, $\chi_4(l,\tau)$ was measured vs time at fixed $l$ for several packing fractions.  In both Refs.~\cite{DauchotSHDPRL2005,KeysAbateNP07}, the dynamic susceptibilities are based on temporal fluctuations of an overlap order parameter; however, the former employs a Gaussian cutoff function while the latter uses a step function.

In this paper we conduct further measurements and analyses of spatially-heterogeneous dynamics (SHD) in a system of air-fluidized beads similar to Ref.~\cite{KeysAbateNP07}.  Our primary extension of Ref.~\cite{KeysAbateNP07} concerns the cutoff function in the overlap order parameter.  First we explore the length-dependence of $\chi_4(l,\tau)$ by varying the cutoff distance $l$.  Second, we remove the arbitrariness in choice of cutoff function by using {\it topological} measures of overlap.  In particular we build upon the concept of persistence, introduced to help characterize the coarsening of spin domains in magnets or of bubbles in foams \cite{BrayEL94, TamPAFoamPRL1997}.  As a topological measure of overlap for particulate systems, we define a normalized, dimensionless persistent area as the fraction of space that remains inside the same Voronoi cell after a given time interval.  We also introduce a second topological measure of configurational overlap in terms of persistent bonds, i.e.\ the fraction of nearest-neighbor pairs that remain neighbors after a given time interval.  While both persistent area and persistent bond order parameters vanish with time, the latter can exhibit a much slower decay if a string of particles follow one another over a long distance.  After discussion of such cutoff functions, we use the resulting four-point susceptibilities both to explore the connection of SHD to local structure and to observe the evolution of SHD on approach to jamming.  Finally we consider possible artifacts due to finite experiment duration.

%=========================================================================================
\section{Methods and Background}

\begin{figure}
\includegraphics[width=3.0in]{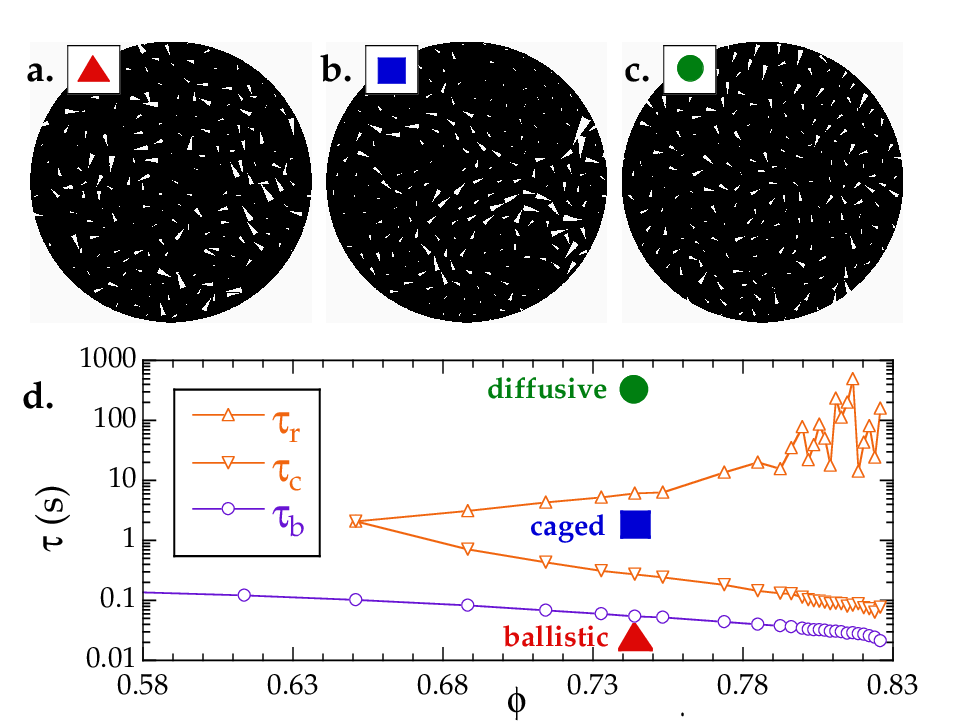}
\caption{(Color online) (a-c) Example average velocity vector fields, $\Delta {\bf r}/\tau$, for time intervals $\tau$ and bead area fractions $\phi$ as specified in (d).  At short times, $\tau<\tau_b$ as in (a), the bead motion is ballistic and there are no spatial correlations.  At very long times, as in (c), the bead motion is diffusive and there are no spatial correlations.  At intermediate times, $\tau_c<\tau<\tau_r$, the bead motion is sub-diffusive with beads remaining temporarily trapped in a cage of fixed neighbors.  Toward the end of this caging regime, as in (b), the bead motion exhibits spatially heterogeneous dynamics in the form of string-like swirls in the average velocity vector field.  This figure was generated from data described in Ref.~\protect{\cite{PreJamming}}, where the bead diameters are 0.873~cm and 0.635~cm, and where precise definitions of time-scales are given in terms of the logarithmic slope of the mean-squared displacement vs time.}
\label{SHDDemo}
\end{figure}

\begin{figure}
\includegraphics[width=3.0in]{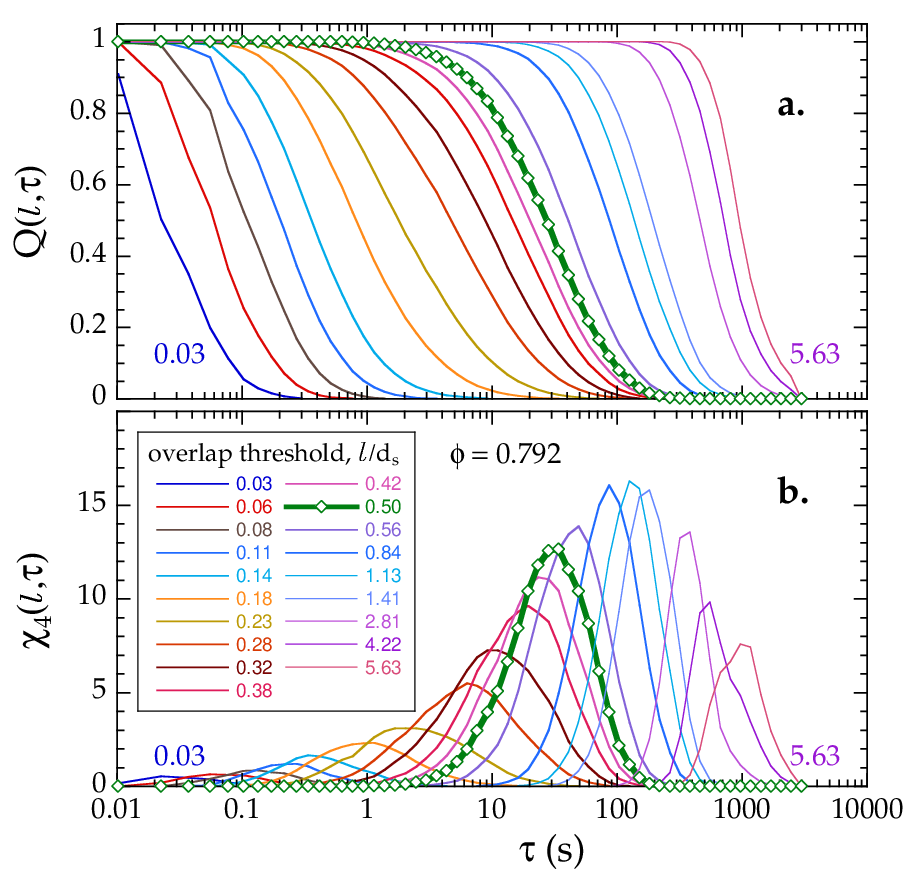}
\caption{(Color online) Time dependence of (a) the average overlap order parameter and (b) the corresponding four-point susceptibilities, defined by Eqs.~(\protect{\ref{soop},\ref{chi4}}) respectively, for air-fluidized beads at an area packing fraction of $\phi=0.792$.  Different curves are for different cutoff lengths, $l$, as labeled.  The thick green curve is for the standard cutoff threshold $l=0.5d_s$ given by half the small-bead diameter.}
\label{Qq}
\end{figure}

\begin{figure}
\includegraphics[width=3.0in]{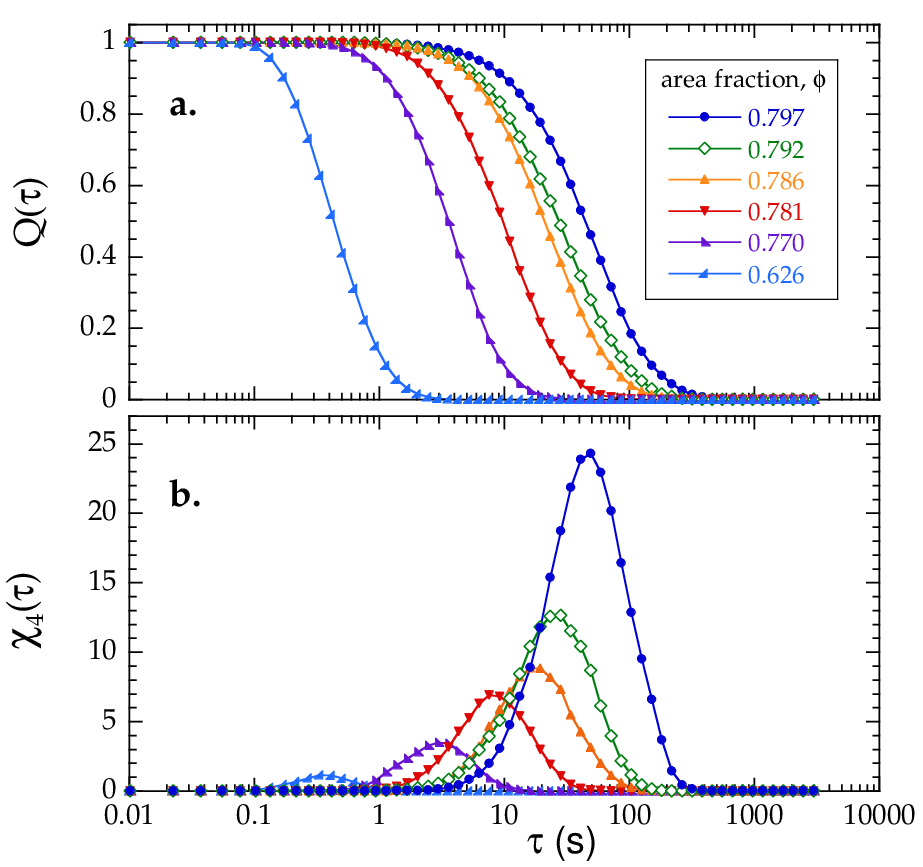}
\caption{(Color online) Time dependence of (a) the average overlap order parameter and (b) the corresponding four-point susceptibilities, for air-fluidized beads at various area fractions, as labeled.  The cutoff threshold is $l=0.5d_s$.}
\label{Qaf}
\end{figure}

The granular system under study here is a quasi-two-dimensional monolayer of spheres that roll without slipping due to a fluidizing upflow of air, as in Refs.~\cite{AbatePills, PreJamming, KeysAbateNP07}.  The spheres are steel with an equal number of $d_s=0.318$~cm and $d_l=0.397$~cm diameter sizes.  This is the same system as in Ref.~\cite{KeysAbateNP07}, though here most run durations are 120~minutes rather than just 20~minutes. The steel beads are confined in a circular flat sieve with a diameter of 17.7~cm and a mesh size of 150~$\mu$m.  Energy is injected by a uniform flow of air through the sieve at a flux which is high enough to randomly drive the balls by turbulence (${\rm Re}\approx 10^3$) without causing levitation.  The flux is increased from 560~cm/s to 700~cm/s as the area fraction is decreased from $\phi=0.8$ to $\phi=0.6$, to ensure that the low density systems are fluidized.  Three feet above the sieve are six incandescent lights aligned in a ring one foot in diameter. At the center of the ring is a 120~Hz Pulnix~6710 video camera with a $644\times484$ array of 8BIT square pixels and a zoom lens. Light reflects specularly off the tops of the beads and is imaged by the camera. Bright spots are tracked so that the positions ${\bf r}_i(t)$ are known for all beads $i=\{1,2,\ldots,N\}$ and for all times $t$ during the entire experiment.  Further details are available in Refs.~\cite{AbatePills, PreJamming}.

By increasing the fraction $\phi$ of area occupied by spheres, at fixed gas flux, this system exhibits a jamming transition at point-J, such that the average bead kinetic energy vanishes linearly on approach to random close packing at about $\phi_c=0.83$~\cite{PreJamming}.  Near this transition, the bead dynamics exhibit the usual sequence of crossovers from ballistic, to sub-diffusive, to diffusive motion as a function of delay time.  Spatially-heterogeneous dynamics occur in the sub-diffusive regime, particularly at delay times such that the beads are breaking out of their cages and beginning to diffuse.  This is illustrated graphically in Fig.~\ref{SHDDemo}(a-c), which show example average velocity vector fields, $\Delta {\bf r}/\tau$, for three different delay times $\tau$; the area fraction and delay time for each image is chosen to highlight motion in each of the three regions of phase space, as specified in Fig.~\ref{SHDDemo}d.  For both very short and very long delay times, corresponding to ballistic and diffusive regimes, the magnitude and direction of the average velocities for neighboring beads are completely uncorrelated.  However, at intermediate delay times, Fig.~\ref{SHDDemo}b, corresponding to subdiffusive motion, a few string-like clusters of neighboring beads move in swirls while other beads move relatively little.  As time evolves, such swirls come and go in different regions of the sample so that the dynamics become completely ergodic.  This is best seen by a video of the average velocity field~\cite{AbateChaos}.

Now we begin the central task of this paper, which is to quantify the spatially-heterogenous nature of the dynamics depicted qualitatively in Fig.~\ref{SHDDemo}. For this, a standard tool is the four-point dynamic susceptibility $\chi_4(l,\tau)$, as reviewed in Ref.~\cite{LacevicJCP2003X4}.  This may be approximated from temporal fluctuations in the instantaneous self-overlap order parameter, defined here as
\begin{equation}
    Q_t(l,\tau) = {1 \over N} \sum_{i=1}^{N}w_i
\label{oop}
\end{equation}
where $N$ is the number of beads and $w_i$ is a cutoff function that equals 1 if the displacement of bead $i$ remains less than $l$ across the whole time interval $t\rightarrow t+\tau$ and that equals 0 otherwise.  Whereas here and in Ref.~\cite{KeysAbateNP07} $w_i$ is a step-function cutoff, in Ref.~\cite{DauchotSHDPRL2005} it is a Gaussian.  In either case the average self-overlap order parameter and the four-point susceptibility are given by moments of $Q_t(l,\tau)$ averaged over all times $t$ as follows:
\begin{eqnarray}
    Q(l,\tau) &=& \langle Q_t(l,\tau) \rangle \label{soop} \\
    \chi_4(l,\tau) &=& N\left[ \langle Q_t(l,\tau)^2 \rangle - \langle Q_t(l,\tau) \rangle^2 \right] \label{chi4}
\end{eqnarray}
Note that $\chi_4(l,\tau)$ is independent of system size because the variance scales as $1/N$.  Also note that $\chi_4$ grows in proportion to the variance, or heterogeneity, in the displacements of different beads across a given time interval.

Example results for the space and time dependence of $Q(l,\tau)$ and $\chi_4(l,\tau)$ are shown in Fig.~\ref{Qq}, for one particular area fraction $\phi=0.792$.  In particular, these two functions are plotted vs delay time $\tau$ for several different values of $l$. By construction $Q(l,\tau)$ decays from one to zero as a function of $\tau$.  For increasing $l$, the observed decay becomes somewhat sharper, but more importantly the location of the decay moves to longer delay times, since the beads take longer to move a larger distance.  By contrast the susceptibility $\chi_4(l,\tau)$ is a peaked function of $\tau$, which vanishes at both early and late times. By comparison of Figs.~\ref{Qq}a-b it is evident that the peak of $\chi_4(l,\tau)$ is located at the characteristic decay time of $Q(l,\tau)$.  Furthermore, the {\it height} of the peak depends on $l$, such that it vanishes for large and small $l$ and becomes a maximum for $l$ on the order of the bead size.  Thus in Ref.~\cite{KeysAbateNP07} the overlap threshold was set equal to the radius of the small beads, $l=0.5d_s$, and the dynamic susceptibility was considered only as a function of time, $\chi_4(\tau)=\chi_4(0.5d_s,\tau)$.

Results for $Q(\tau)$ and $\chi_4(\tau)$ vs $\tau$ are displayed in Figs.~\ref{Qaf}a-b for a sequence of area fractions.  This reproduces Fig.~3c of Ref.~\cite{KeysAbateNP07}, with the identical system but with a longer run duration.  As seen before, the location $\tau_4^*$ of the peak in $\chi_4(\tau)$ moves to longer times on approach to jamming.  Furthermore, the peak height $\chi_4^*$ increases concurrently.  Thus the dynamics become not only slower but more spatially heterogeneous on approach to jamming.  The size of the spatial heterogeneities represents a dynamical correlation length, $\xi$, which diverges at jamming; such behavior was first reported in \cite{KeysAbateNP07}, and will be discussed further in Section IV.  Note that for all area fractions, the value of the overlap order parameter is nearly constant, $Q^*=0.48\pm0.06$, when the susceptibility reaches its peak.

The dynamical correlation length may be deduced from the value of the self-overlap order parameter and the dynamic susceptibility at its peak, using the following physical picture.  For a system of $N$ total beads, we wish to know the typical number $M$ of heterogeneities and more importantly the typical number $n$ of beads in each of these fast-moving domains.  Since the contribution to the self-overlap order parameter is $Q_0=0$ from the $nM$ highly-mobile beads inside domains, and is $Q_1=1$ from the other $N-nM$ less-mobile beads, the average self-overlap order parameter at $\tau_4^*$ is $Q^* \equiv \langle Q_t(\tau_4^*)\rangle = [Q_0(nM)+Q_1(N-nM)] / N = 1-nM/N$.  However the instantaneous value of $Q_t(\tau_4^*)$ varies with time $t$ because of counting statistics $\Delta M=\sqrt{M}$ in the number of heterogeneous domains actually present at any instant.  Therefore, the peak dynamic susceptibility is $\chi_4^* \equiv N(\Delta Q^*)^2=n^2M/N$.  By eliminating $M/N$ in these expressions for $Q^*$ and $\chi_4^*$, the typical number of particles in each fast-moving dynamic heterogeneity is found to be
\begin{equation}
    n_4={\chi_4^* \over 1-Q^*}.
\label{nq}
\end{equation}
Since $Q^*$ is nearly constant, as observed in Fig.~\ref{Qaf}, the number of beads per domain is directly proportional to the peak susceptibility $\chi_4^*$.  If the domains are compact, then the dynamical correlation length scales as $\xi/d_s \sim n^{1/d}$, where $d_s$ is the bead diameter and $d$ is dimensionality \cite{DauchotSHDPRL2005, KeysAbateNP07}.  If the domains are strings, then the dynamical correlation length is $\xi/d_s\sim n$.  If fluctuations $\Delta n$ in domain size are also included, then the left-hand side of Eq.~(\ref{nq}) is slightly modified to $n[1+(\Delta n/n)^2]$.

To our knowledge, the counting arguments leading to Eq.~(\ref{nq}) have not been previously published.  While it seems common knowledge that the peak susceptibility depends upon domain size, we have found no arguments or calculations in prior literature to support or quantify the relation.

%=========================================================================================
\section{Topological Persistence}

%=========================================================================================

The dynamic susceptibility $\chi_4(l,\tau)$ discussed in the previous section relies upon a somewhat arbitrary choice for a cutoff function, $w$, in Eq.~(\ref{oop}).  Furthermore, since $\tau$ is the more interesting parameter, it also relies upon a particular choice of cutoff length, $l$, so that spatially-heterogeneous dynamics may be studied as a function of delay time.  In this section we propose two alternative dynamic susceptibilities, intended for the same purpose, in which the role of the cutoff function and cutoff length are determined uniquely by topological considerations.  Both are based upon the Voronoi tessellation constructed from the particle positions at each instant of time.

\subsection{Persistent area}

\begin{figure*}
\includegraphics[width=6.5in]{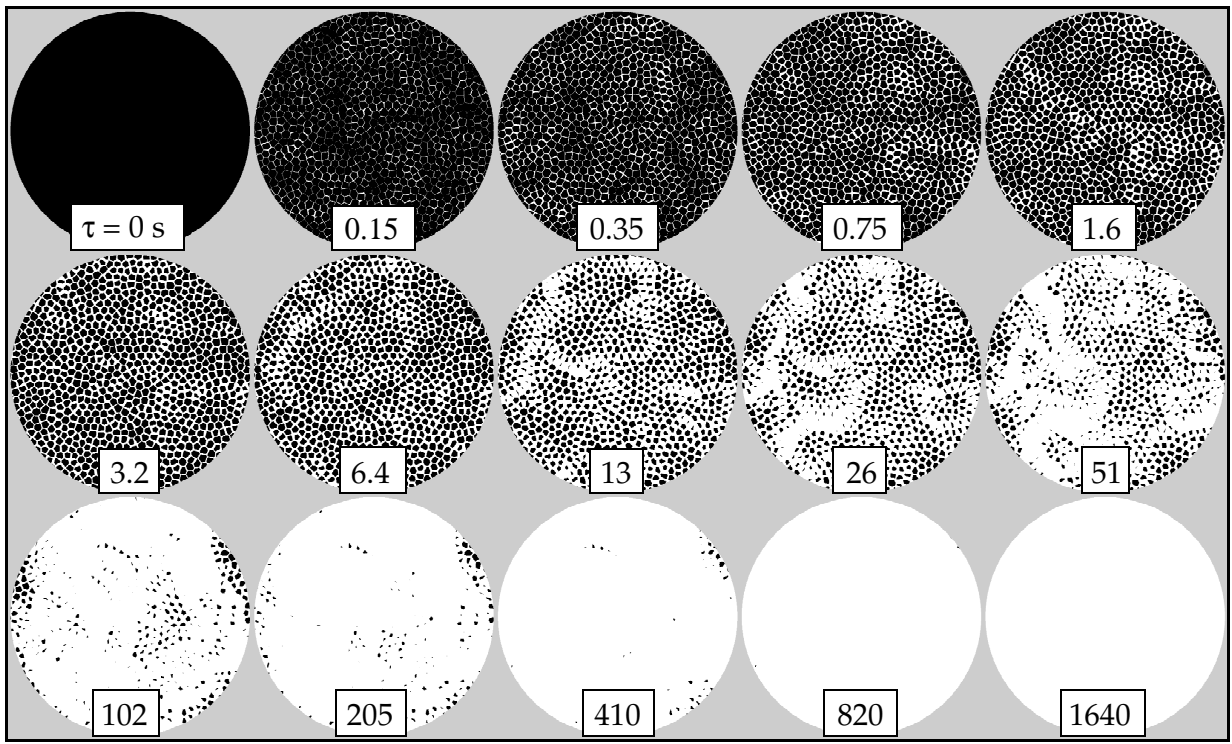}
\caption{Persistent area images vs delay time, $\tau$, as labeled, for air-fluidized beads at an area packing fraction of $\phi=0.792$.  The shrinking black regions represent persistent area, which has remained inside the same Voronoi cell since zero delay time.  Each graphic represent the central 7.5~cm diameter region of the entire 17.7~cm diameter sample.  The dynamics are most spatially heterogeneous at delay time $\tau_A^*=34$~s where the susceptibility $\chi_A(\tau)$ reaches a maximum.}
\label{PADemo}
\end{figure*}

\begin{figure}
\includegraphics[width=3.0in]{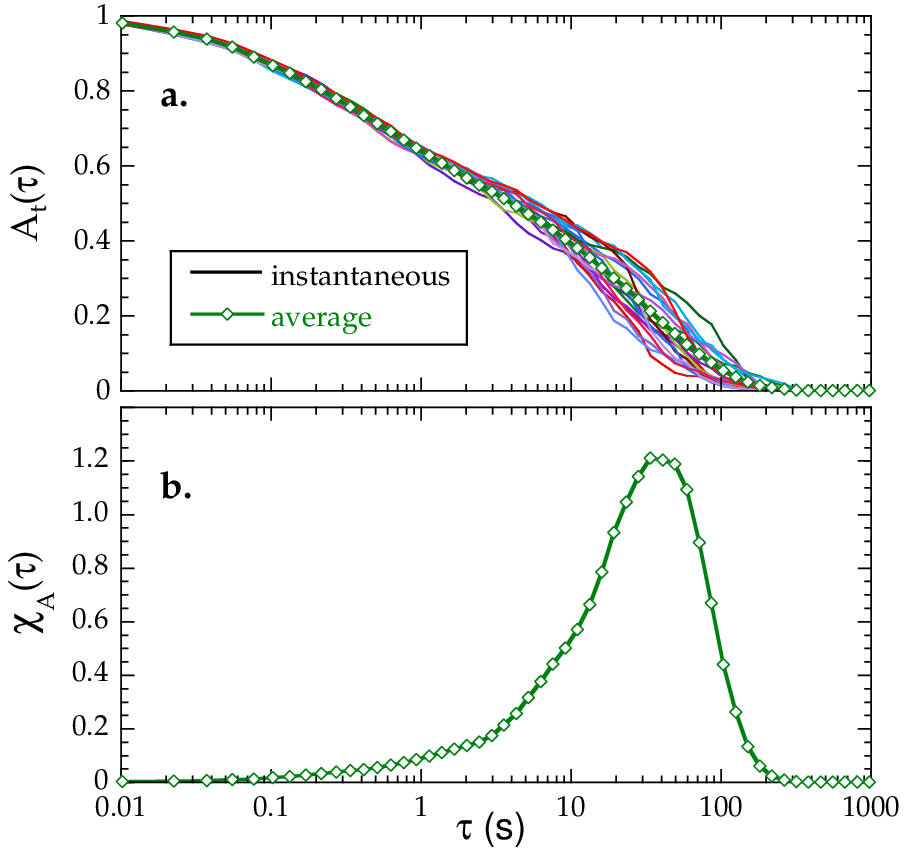}
\caption{(Color online) Time dependence of (a) the instantaneous and average persistent areas, defined by Eqs.~(\protect{\ref{ipa},\ref{apa}}), respectively, and (b) the corresponding dynamic susceptibility, defined by Eq.~(\protect{\ref{chiA}}), for air-fluidized beads at an area packing fraction of $\phi=0.792$.  The starting times $t$ for the instantaneous persistent areas are separated by 180~s, and thus represent statistically independent initial configurations.}
\label{Amv}
\end{figure}

\begin{figure}
\includegraphics[width=3.0in]{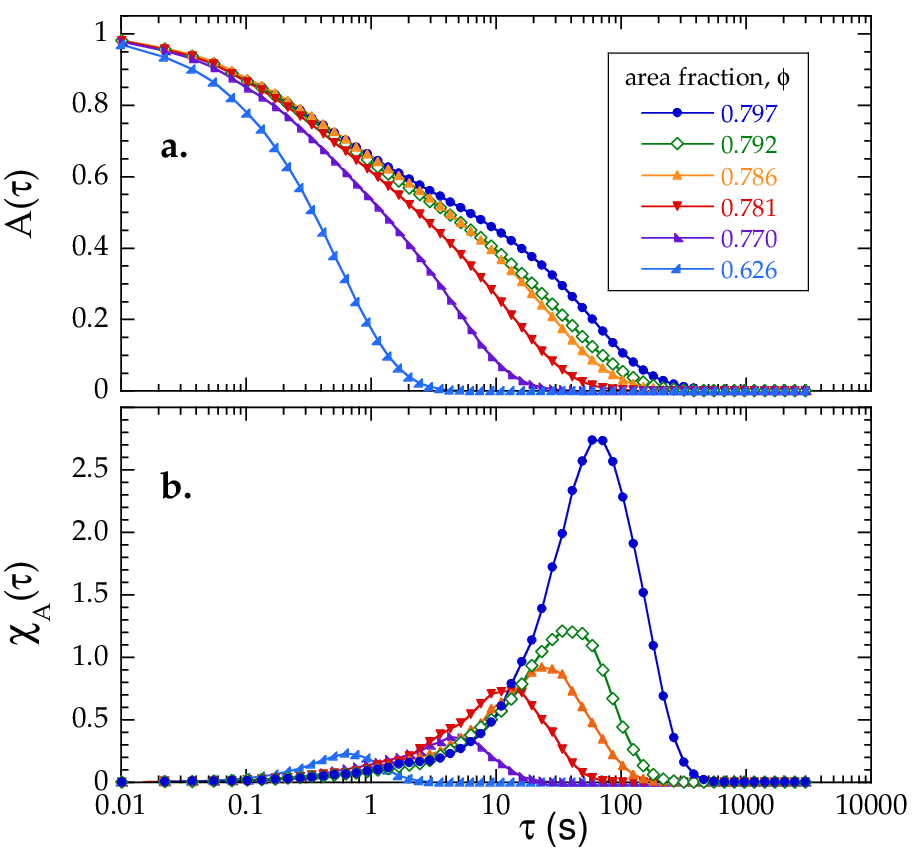}
\caption{(Color online) Time dependence of (a) the average persistent area, and (b) the corresponding dynamic susceptibility, for air-fluidized beads at different area fractions as labeled.}\label{Aaf}
\end{figure}

An order parameter for quantifying the degree of overlap between particle configurations may be constructed by considering the fraction of the sample which remains inside the same Voronoi cell across a time interval $t\rightarrow t+\tau$.  In particular, we define the instantaneous persistent volume as
\begin{equation}
    A_t(\tau) = {1\over V}\int a_t({\bf r},\tau){\rm d}V
\label{ipa}
\end{equation}
where $V$ is the sample volume, $a_t({\bf r},\tau)=1$ if the point $\bf r$ remains inside the {\it same} Voronoi cell across the whole time interval $t\rightarrow t+\tau$, and $a_t({\bf r},\tau)=0$ if the point $\bf r$ becomes enclosed by another Voronoi cell.  Since our experiments are in two-dimensions, we shall refer to Eq.~(\ref{ipa}) as the instantaneous persistent area.  By construction, it decays monotonically from one to zero as the beads move a distance comparable to their size.  Thus the persistent area is similar to the previous self-overlap order parameter $Q_t(l,\tau)$ with a cutoff length $l$ set by bead size.

A graphic illustration of the contribution $a_t({\bf r},\tau)$ of each point in the sample to the average over space given by Eq.~(\ref{ipa}) is shown in Fig.~\ref{PADemo} for a sequence of increasing delay times $\tau$.  Pixels are colored black for $a_{t,\tau}({\bf r})=1$ and are colored white otherwise; therefore, the sample is all black for $\tau=0$ and it becomes progressively white for increasing $\tau$.  As illustrated in Fig.~\ref{PADemo}, white regions first appear at short $\tau$ near the boundary of the initial Voronoi cells.  At larger $\tau$ the white regions thicken as the persistent black regions at the cell cores shrink.  Eventually each black spot vanishes.  The graphics here can be compared with results for a coarsening foam in Ref.~\cite{TamPAFoamPRL1997}, where domains are defined physically by actual bubbles rather than by a Voronoi construction.  For dynamics that are spatially uniform, the persistent area vanishes uniformly at the same rate throughout the whole sample, with only random uncorrelated variation between neighboring regions.  This is not the case in Fig.~\ref{PADemo}, where the spatially-heterogeneous nature of the dynamics is evident in the long swaths of white caused by chains of fast-moving beads.  This is most pronounced for delay times between about 5 and 50~s, and corresponds closely with structure in the average velocity field seen for example in Fig.~\ref{SHDDemo}.

The average persistent area and a dynamical susceptibility quantifying spatial heterogeneity may now be defined, in analogy with the previous section, by moments of $A_t(\tau)$ averaged over all times $t$:
\begin{eqnarray}
    A(\tau) &=& \langle A_t(\tau) \rangle \label{apa} \\
    \chi_A(\tau) &=& N\left[ \langle A_t(\tau)^2 \rangle - \langle A_t(\tau) \rangle^2 \right] \label{chiA}
\end{eqnarray}
The variance is multiplied by the number $N$ of beads in the sample, so that $\chi_A(\tau)$ is independent of system size.  As an example, the decays of $A_t(\tau)$ vs $\tau$ plotted in Fig.~\ref{Amv}a display considerable variation for different choices of starting time $t$.  The resulting susceptibility, $\chi_A(\tau)$, plotted underneath in Fig.~\ref{Amv}b, displays a peak as expected near the characteristic decay time for the average persistent area.  The height of this peak, $\chi_A^*$, and the corresponding average persistent area $A^*$, give the typical number of beads in the dynamic heterogeneities as
\begin{equation}
    n_A={\chi_A^* \over (A_1-A_0)(A_1-A^*)},
\label{na}
\end{equation}
by the same arguments giving Eq.~(\ref{nq}).  Here, $A_0$ is the contribution to the persistent area by the beads in the fast-moving  heterogeneities and $A_1$ is the contribution from the remaining less mobile beads.  Based on graphics like Fig.~\ref{PADemo}, we estimate $A_0\approx0$ and $A_1\approx3/8$.  To remove this uncertainty, an alternative approach might be to enforce $A_0=0$ and $A_1=1$ by defining an overlap order parameter such that the contribution from each bead is a step function that drops from one to zero when all its persistent area vanishes.

The behavior of $A(\tau)$ and $\chi_A(\tau)$ are displayed in Fig.~\ref{Aaf} for different bead packing fractions, $\phi$, using the same sequence of values and the same symbol/color codes as in Fig.~\ref{Qaf}.  The decay time of $A(\tau)$ and the peak position $\tau_A^*$ of $\chi_A(\tau)$ coincide and become longer for increasing $\phi$ on approach to jamming.  For all area fractions, the value of the persistent area is nearly constant, $A^*=0.24\pm0.04$, when the susceptibility reaches its peak.  The peak height $\chi_A^*$ also increases as the dynamics slow down, indicative of a diverging dynamical correlation length.  This behavior strongly parallels that in Fig.~\ref{Qaf} based on an overlap order parameter with a particular step-function cutoff.  In fact, the ratio of peak heights is nearly constant for all area fractions: $\langle \chi_A^*/\chi_4^* \rangle = 0.10\pm0.02$.  The only striking qualitative difference is that $A(\tau)$ exhibits a two-step decay at high $\phi$, whereas $Q(\tau)$ does not.  This is because the persistent area begins to decay at very short delay times, when the beads experience small-displacement ballistic motion, rattling within a ``cage'' of unchanging nearest neighbors.  The shoulder in $A(\tau)$ corresponds to the crossover from ballistic to subdiffusive motion observed in the mean-squared displacment.  For a typical cutoff $l$ comparable to bead size, this so-called $\beta$ relaxation is not detected by the traditional overlap order parameter $Q(\tau)$, which decays only due to the cage-breakout $\alpha$-relaxation process.  Furthermore, no shoulder is evident in $Q(l,\tau)$ for any choice of $l$ in Fig.~\ref{Qq}.  Besides being naturally and uniquely defined by topology, the persistent area order parameter thus has the added advantage of being able to capture the two-step $\beta$ and $\alpha$ relaxation processes characteristic of glass-forming systems.

%=========================================================================================
\subsection{Persistent bond}

\begin{figure*}
\includegraphics[width=6.5in]{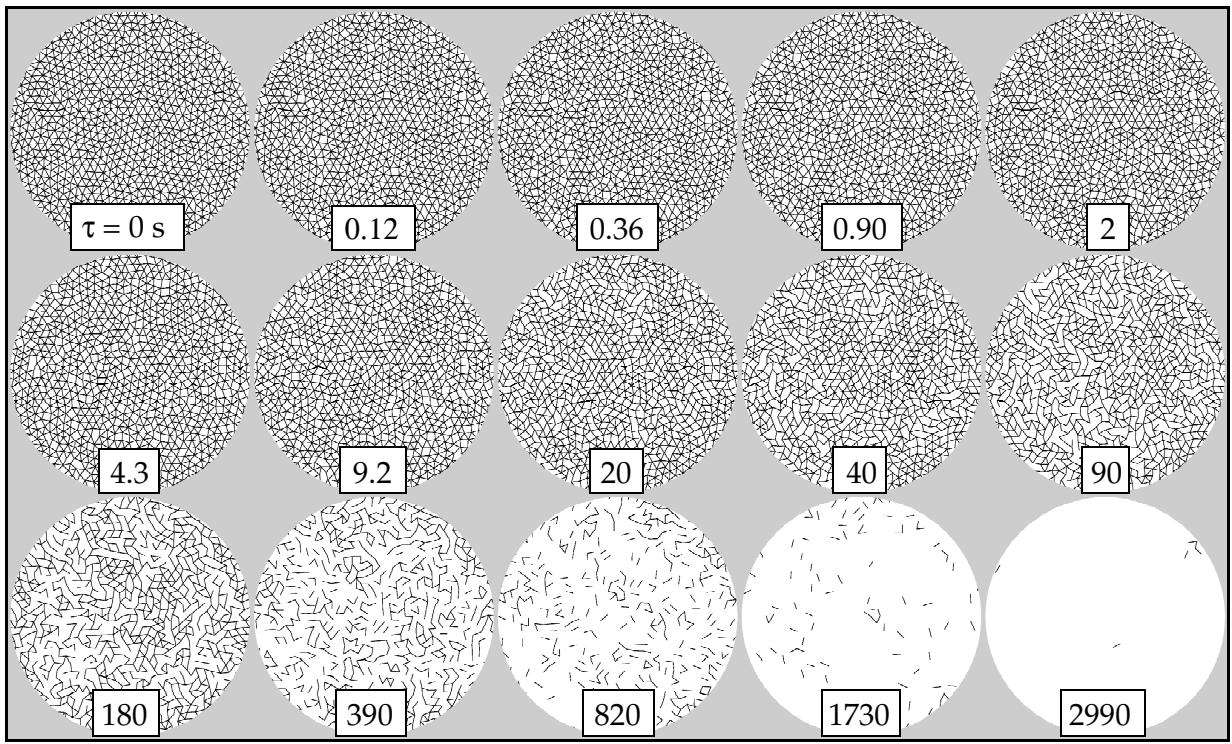}
\caption{(Color online) Persistent bond images vs delay time, $\tau$, as labeled, for air-fluidized beads at an area packing fraction of $\phi=0.792$. Each graphic represent the central 7.5~cm diameter region of the entire 17.7~cm diameter sample.  The dynamics are most spatially heterogeneous at delay time $\tau_B^*=180$~s where the susceptibility $\chi_B(\tau)$ reaches a maximum}
\label{PBDemo}
\end{figure*}

\begin{figure}
\includegraphics[width=3.0in]{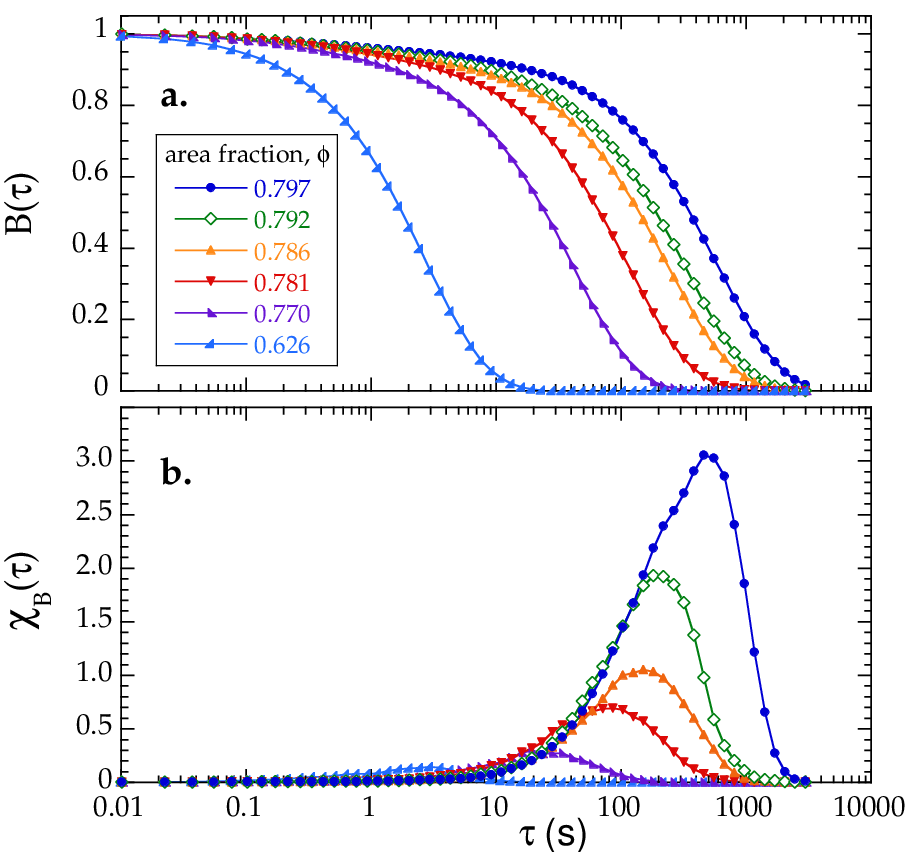}
\caption{(Color online) Time dependence of (a) the average persistent bond, and (b) the corresponding dynamic susceptibility, for air-fluidized beads at different area fractions as labeled.}
\label{Baf}
\end{figure}

A second order parameter may be defined naturally from the set of nearest neighbors, or ``bonds'', specified by the Delaunay triangulation dual to the Voronoi construction.  In particular, we define the instantaneous persistent bond number $B_t(\tau)$ as the fraction of all bonds at time $t$ that remain unbroken across the interval $t\rightarrow t+\tau$.  The progressive breaking of bonds throughout the sample is illustrated in Fig.~\ref{PBDemo}, at the same area fraction as in the persistent area fraction diagrams of Fig.~\ref{PADemo}.  By comparison, the persistent bond number decays more slowly, due to neighboring beads that move a large distance while remaining next to one another.  Also by comparison, the string-like swirls of the dynamic spatial heterogeneities are less evident by casual inspection of the persistent bond diagrams than for the persistent area diagrams.  Nevertheless, bond lifetimes and the spatial heterogeneity of broken bonds have been used to characterize simulations of glassy systems~\cite{YamamotoPRE1998AtomicStrings}.

The average persistent bond number, the related dynamic susceptibility, and the size of the dynamic heterogeneities, respectively, are given in precise analogy with the previous sections as follows:
\begin{eqnarray}
    B(\tau) &=& \langle B_t(\tau) \rangle \label{apb} \\
    \chi_B(\tau) &=& N\left[ \langle B_t(\tau)^2 \rangle - \langle B_t(\tau) \rangle^2 \right] \label{chiB} \\
    n_B &=& {\chi_B^* \over (B_1-B_0)(B_1-B^*)} \label{nb}
\end{eqnarray}
The trends in this order parameter and susceptibility are displayed vs $\tau$ in Figs.~\ref{Baf}a-b for the same sequence of area fractions as in the analogous plots based on self-overlap and persistent area.  Here, as before, the order parameter decays more slowly and the peak height increases as the area fraction is increased towards jamming.  Similar to $Q(\tau)$, but in contrast with $A(\tau)$, the average persistent bond number $B(\tau)$ exhibits a one-step decay.  By inspection, we estimate $B_0\approx1/3$ and $B_1\approx2/3$.  Note that for all area fractions, the value of the persistent bond is nearly constant, $B^*=0.46\pm0.08$, when the susceptibility reaches its peak.  Also, the ratio of peak susceptibility heights is nearly constant: $\langle \chi_B^*/\chi_4^* \rangle = 0.13\pm0.05$.

%=========================================================================================
\section{Comparisons}

\begin{figure*}
\includegraphics[width=6.5in]{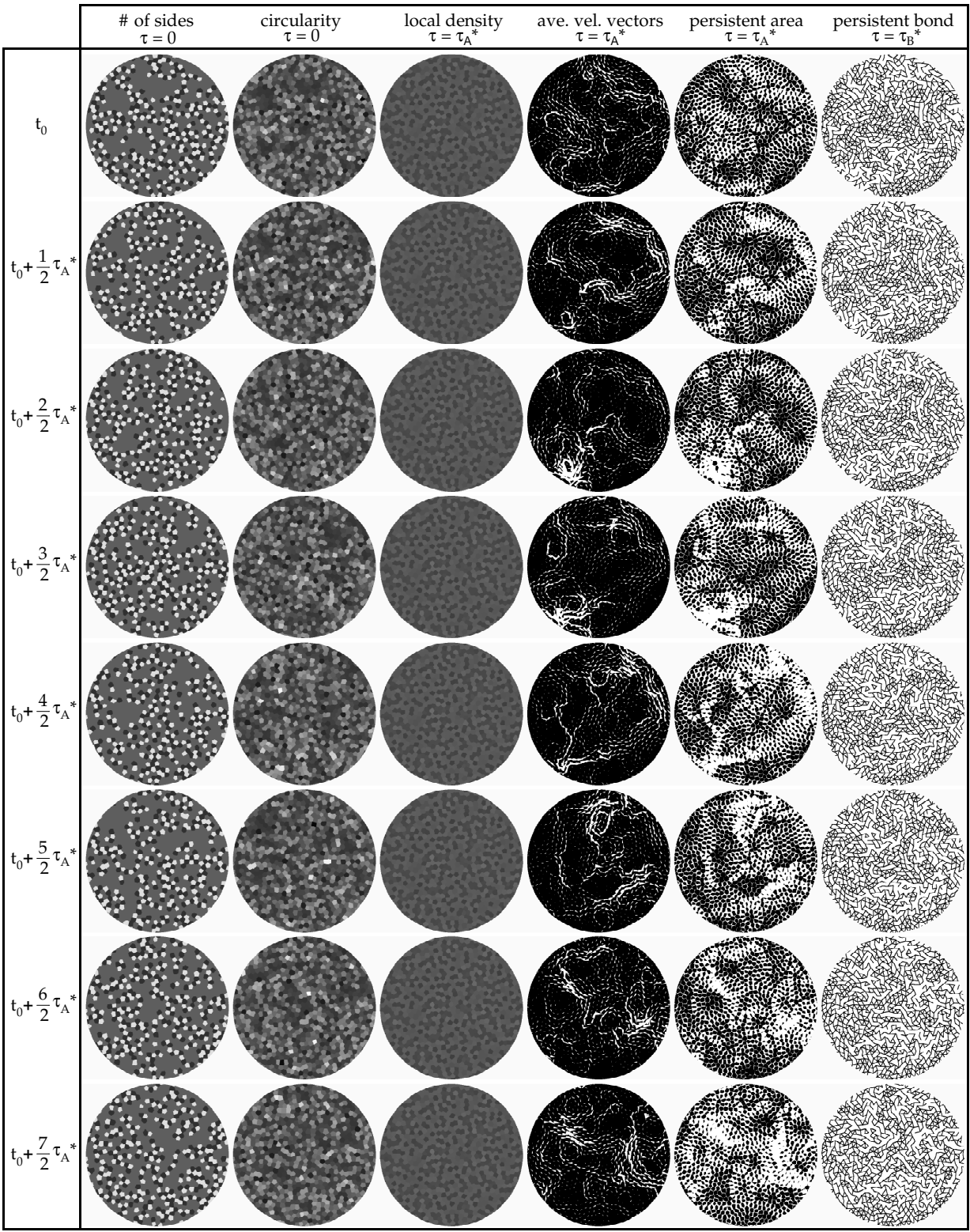}
\caption{(Color online) Graphic illustration of three measures of structure and three measures of dynamics, at eight instants of time. Column-1: Voronoi cells shaded according to their number of sides $Z$; cells with many sides are darker than cells with few. Column-2: Voronoi cells shaded according to their circularity $\zeta$, defined by perimeter squared divided by $4\pi$ times area; more circular cells are darker than less circular. Column-3: Voronoi cells shaded according to local density, proportional to the reciprocal of their area and averaged over the time interval $\tau_A^*$; cells with low local density are colored darker than cells with high local density.  The standard deviation of density values equals about one-tenth the average density.  Columns 4-6: average velocity vector fields, persistent areas, and persistent bonds, respectively, at delay times as labeled.}\label{ZCDVAB}
\end{figure*}

\begin{figure*}
\includegraphics[width=6.5in]{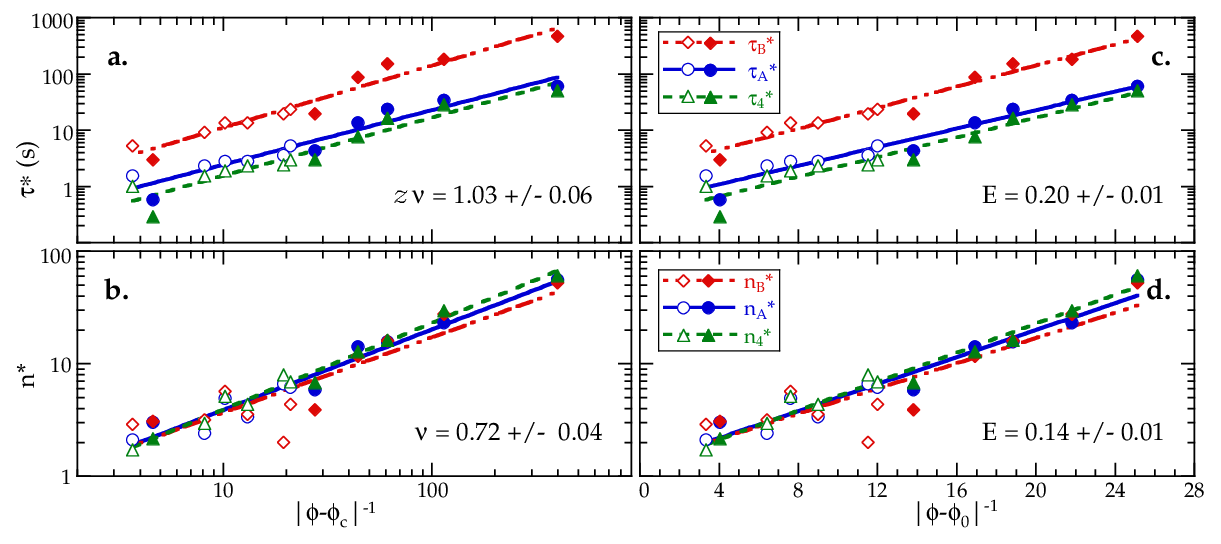}
\caption{(Color online) Growth with increasing area fraction of (a,c) times at which the three dynamic susceptibilities $\chi_Q(\tau)$, $\chi_A(\tau)$, and $\chi_B(\tau)$ become maximum and (b,d) the domain sizes for the spatial dynamic heterogeneities as deduced from Eqs.~(\protect{\ref{nq}, \ref{na}, \ref{nb}}).  Best fits to power law divergence as $|\phi-\phi_c|^{-\lambda}$ yield a critical density of $\phi_c=0.79 \pm 0.01$ and exponents as labeled.  Best fits to exponential divergence as $\exp(E/|\phi-\phi_0|)$ yield a Vogel-Thammann-Fulcher critical density at $\phi_0=0.83 \pm 0.01$ and energies as labeled. Solid symbols are for the 120~minute runs featured in previous plots; open symbols are for 20~minute runs.}\label{TimeLeng}
\end{figure*}

Three dynamic susceptibilities, $\chi_4(\tau)$, $\chi_A(\tau)$, and $\chi_B(\tau)$, have now been discussed for characterizing spatial dynamic  heterogeneities in terms of the variance of order parameters based respectively on self-overlap, persistent area, and persistent bond.  The nature of these order parameters, and their possible relation with local structure, are compared in Fig.~\ref{ZCDVAB} by snapshots of the system at eight different instants in times.  Here the time increment is chosen as $\tau_A^*/2$, half the time delay at which $\chi_A(\tau)$ reaches its peak, so that there is a reasonable balance of continuity and evolution between successive snapshots.  While each row represents a different time, the first three columns represent local structure and the last three represent local dynamics.  In terms of structure, Voronoi cells are shaded according to (a) their number of sides; (b) their circularity shape factor, equal to perimeter-squared divided by $4\pi$ times area \cite{Moucka, ShattuckPRL06, PreJamming}; and (c) the reciprocal of their area, as a measure of the local density.  In terms of dynamics, the average velocity vectors and the persistent areas are depicted at $\tau_A^*$ and the persistent bonds are depicted at $\tau_B^*$, so that the spatial dynamic heterogeneities are maximally emphasized.  By inspection, we note the following points.  First, the spatial structure of the dynamics revealed by the average velocity vectors and the persistent areas is clearly related.  Both measures give a clear feeling for string-like swirls of neighboring beads that is intermittently excited against a background of less mobile beads.  By contrast the persistent bonds give no such feeling of motion.  And while there are more broken bonds near the fast moving domain, the correspondence isn't one-to-one because neighboring mobile beads may remain nearest neighbors.  Second, there is no apparent correlation between any of the measures of structure and the location of the spatial dynamic heterogeneities.  It is not evident how to predict when or where a dynamic heterogeneity will occur based on usual measures of local structure.

Next we compare the characteristic times and domain sizes for dynamic heterogeneities, as deduced from the three order parameters and susceptibilities.  The growth of these scales as a function of increasing bead area fraction are displayed in Fig.~\ref{TimeLeng}.  Note that the time scales are different for the three susceptibilities.  As expected, $\tau_B^*$ is the largest while $\tau_A^*$ is only slightly larger than $\tau_4^*$.  However all three times appear to grow in proportion to one another, suggesting that they probe similar physics.  Further reinforcing this point, the domain sizes are indistinguishable for the three susceptibilities.  Thus the overlap order parameter with discretionary choice of step-function cutoff at $l=0.5d_s$ is not as arbitrary as it may seem, since it reveals the same picture of spatially heterogeneous dynamics as the persistent areas and bonds, which are uniquely defined by topology.

Next we consider the functional form of the growth of dynamical heterogeneities on approach to jamming, exactly per Ref.~\cite{KeysAbateNP07}.  Thus in Figs.~\ref{TimeLeng}a-b, data for the peak times and domain sizes are shown on log-log plots vs $|\phi-\phi_c|^{-1}$.  With a single choice of $\phi_c=0.79\pm0.02$, the data can all be well-fit to a power of $|\phi-\phi_c|^{-1}$ where the exponent is $\nu=0.72\pm0.04$ for the domain sizes and is $z\nu=1.03\pm0.06$ for the peak times.  Such power-law divergences on approach to $\phi_c$ below random-close packing are consistent with a mode-coupling theory, where the mode-coupling temperature is above the glass-transition temperature.  The value of the correlation length exponent $\nu$ is consistent with prior simulation results based on finite-size scaling analysis of random close packing fraction \cite{OHernPRE03}, on the size of the disturbance away from a perturbation \cite{ReicchardtPRL05}, and on scaling analysis of stress and strain rate \cite{Teitel07}.  Slightly smaller exponents for the dynamical correlation length have been predicted for the approach $\phi_c$ from above \cite{WyartEL05, SilbertPRL05, JenEL06}.  The observed time scale exponent is less than that for simulation results based on stress relaxation \cite{DurianPRL95, DurianPRE97}, for the same model as in Ref.~\cite{Teitel07}.  However the power-law description is not unique.  As for the glass transition in molecular liquids, the growth of characteristic time and length scales can also be well-fit to Vogel-Tammann-Fulcher (VTF) equation, $\exp(E/|\phi-\phi_0|)$.  Thus in Figs.~\ref{TimeLeng}c-d, data are shown on semi-log plots vs $|\phi-\phi_0|^{-1}$.  With a single choice of $\phi_0=0.83\pm0.01$, which corresponds to random close packing, the data follow the VFT equation where $E=0.20\pm0.01$ for the peak times and $E=0.14\pm0.01$ for the domain sizes.  Altogether, as reported in Ref.~\cite{KeysAbateNP07}, the growth of dynamical time and length scales on approach to point-J in the system of air-fluidized beads happens in close quantitative analogy to the glass transition in liquids.

\begin{figure}
\includegraphics[width=3.0in]{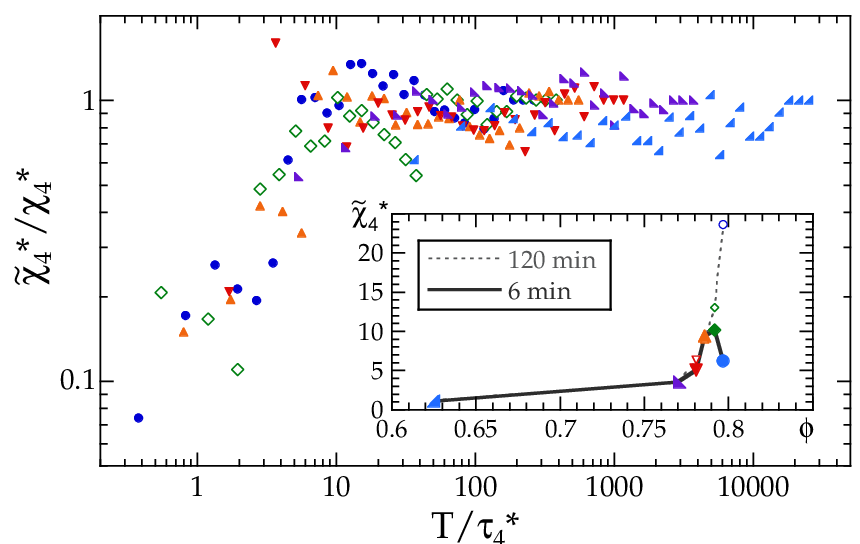}
\caption{(Color online)
Peak height $\tilde\chi_4^*$ vs observation time $T$, for different area fractions as labeled.  Heights and times are normalized, respectively, by the actual long-duration values $\chi_4^*$ and the peak times $\tau_4^*$.  To obtain good configurational averaging and accurate experimental results, the duration of the observation time must be about a decade longer than the decay time of the overlap order parameter.  Inset: Apparent peak height vs area fraction for short and long observation times, as labeled.  Symbols in the main plot denote different area fractions, as specified by corresponding symbols in the inset.}\label{hvT}
\end{figure}

Before closing we make one final comparison of the four-point dynamic susceptibility for runs of different duration.  In particular, we compute peak heights $\tilde\chi_4^*$ for subsets of duration $T$ of the full 120~minute runs and normalize by the long-duration value $\chi_4^*$.  Results are plotted in Fig.~\ref{hvT} vs duration time, normalized by the long-duration peak time $\tau_4^*$, for all area fractions considered above.  Note that the normalization causes all data to collapse, even though peak heights and times grow on approach to jamming.  Also note that the long-duration value of the peak height is attained only if the run is about ten times longer than the peak time $\tau_4^*$.  Intuitively, if a run is not very long compared to the decay time of the overlap order parameter, then too few configurations are sampled and the variance is underestimated.  This effect will limit the proximity to which point-J may be approached.  For example, if run duration is held fixed and the area fraction is gradually increased, then the peak height and the corresponding dynamical correlation length will initially grow but will eventually appear to decrease as $\tau_4^*$ grows beyond about one-tenth of the run duration.  This can be seen in the inset of Fig.~\ref{hvT}.  To eliminate such erroneous finite-time artifacts, the experimental observation time must be at least a decade longer than the time it takes for the overlap order parameter to vanish.  Conceivably, an alternative might be to take a smaller cutoff in the self-overlap order parameter or to alter the proportionality constant relating susceptibility and domain size according to the experimental conditions.

%=========================================================================================
\section{Conclusion}

Spatially-heterogeneous dynamics may be characterized successfully using four-point susceptibilities based on many choices for the configurational order parameter.  For a self-overlap order parameter with the seemingly {\it ad-hoc} choice of a step-function cutoff at $d_s/2$, the characteristic dynamical correlation length for the size of intermittent mobile domains is indistinguishable from those based on persistent areas and bonds in a unique topological description.  This agreement requires a physical picture for how to deduce domain size from the peak of the susceptibility, as in Eqs.~(\ref{nq},\ref{na},\ref{nb}).  Nevertheless, some differences still exist in the detailed form of the three susceptibilities.  For example the peak times all differ, though they always remain in constant proportion to one another.  At early times, below the peak, the persistent area is the first to begin its decay since self-overlap and persistent bonds are insensitive to the small ballistic motion of grains as they rattle in a cages of fixed nearest neighbors.  At late times, past the peak, the persistent bond is the last to decay since a few neighboring beads remain in close contact as they diffuse through the sample.  Thus, while all three susceptibilities give the same picture of the spatially heterogeneous dynamics, they offer different possibilities for characterizing short and long time dynamics.  As applied to the change in behavior with increasing bead packing fraction, all three give characteristic time and length scales for the dynamical heterogeneities that appear to diverge in accord with simulation results for supercooled liquids and for dense athermal systems of soft repulsive particles.

\begin{acknowledgments}
We thank Maksym Artomov, Sharon Glotzer, Aaron Keys, Andrea Liu, and Steve Teitel for helpful discussions. Our work was supported by the National Science Foundation through grants DMR-0514705 and DMR-0704147.
\end{acknowledgments}

\bibliography{topology_refs}

\end{document}